\documentclass[aps,pra,twocolumn,superscriptaddress,longbibliography]{revtex4-2}

\usepackage{graphicx}
\usepackage[usenames]{xcolor}
\usepackage{dcolumn}
\usepackage{soul}
\usepackage{cancel}
\usepackage{xcolor}
\usepackage{bm}
\usepackage{amssymb, amsmath}
\usepackage{physics}
\usepackage{verbatim}
\usepackage{amsmath}
\usepackage[normalem]{ulem}
\usepackage{lineno}
\usepackage{float}
\usepackage{comment}
\usepackage[utf8]{inputenc}
\usepackage[colorlinks,urlcolor=blue,citecolor=blue,linkcolor=blue]{hyperref}
\usepackage{cleveref}

\usepackage{times}

\begin{document}

\title{Robust iSWAP gates for semiconductor spin qubits with local driving}

\author{Qi-Pei Liu}
\affiliation{Key Laboratory of Atomic and Subatomic Structure and Quantum Control (Ministry of Education),\\
Guangdong Basic Research Center of Excellence for Structure and Fundamental Interactions of Matter,\\
and School of Physics, South China Normal University, Guangzhou 510006, China}

\author{ Zheng-Yuan Xue}
\email{zyxue83@163.com}

\affiliation{Key Laboratory of Atomic and Subatomic Structure and Quantum Control (Ministry of Education),\\
Guangdong Basic Research Center of Excellence for Structure and Fundamental Interactions of Matter,\\
and School of Physics, South China Normal University, Guangzhou 510006, China}
\affiliation{Guangdong Provincial Key Laboratory of Quantum Engineering and Quantum Materials,\\
Guangdong-Hong Kong Joint Laboratory of Quantum Matter, and Frontier Research Institute for Physics,\\
South China Normal University, Guangzhou 510006, China}
\affiliation{Hefei National Laboratory, Hefei 230088, China}

\date{\today}

\begin{abstract}
Scalable quantum computation demands high-fidelity two-qubit gates. However, decoherence and control errors are inevitable, which can decrease the quality of implemented quantum operations.
We propose a robust iSWAP gate protocol for semiconductor spin qubits, which is a promising platform for scalable quantum computing. Our scheme uses only local microwave drives on conventional exchange-coupled spin qubits.
This approach simultaneously addresses two critical challenges on semiconductor quantum computing: it suppresses low-frequency noise via continuous dynamical decoupling, and it circumvents the control difficulties associated with the ac modulation of the exchange interaction.
We further develop a composite pulse sequence to remove drive-strength constraints and a dynamically corrected method to provide first-order immunity to microwave amplitude errors.
Numerical simulations confirm that our scheme can achieve fidelity above the fault-tolerance threshold under current experimental conditions, offering a building block for practical quantum processors.
\end{abstract}


\maketitle

\section{Introduction}
Spin qubits in semiconductor quantum dots are promising candidates for scalable quantum computing, due to its compatibility with semiconductor manufacturing for high-density integration~\cite{loss1998quantum,zwanenburg2013silicon,chatterjee2021semiconductor,RevModPhys.95.025003}. Recent advances demonstrate single-qubit fidelities exceeding $99.9\%$~\cite{veldhorst2014addressable,yoneda2018quantum} and two-qubit gate fidelities exceeding the fault-tolerance threshold of $99\%$ ~\cite{noiri2022fast,xue2022quantum,mills2022two,fowler2012surface}, paving the way for quantum error correction, which is a vital step towards scalable quantum computation.
The implementation of two-qubit gates is usually more demanding, as they require delicate control over the coupling between qubits. Current high-fidelity two-qubit gates in semiconductor spin qubits are predominantly controlled rotation and controlled phase (CPhase) gates~\cite{noiri2022fast,xue2022quantum, mills2022two}, which are locally equivalent up to single-qubit rotations. Beyond these gates, the Clifford group encompasses two distinct classes of non-trivial two-qubit gates: SWAP-type and iSWAP-type gates. These gates can substantially reduce circuit depth compared to relying solely on CPhase gates~\cite{abrams2020implementation}. In particular, iSWAP gates enable fermionic-swap operations crucial for quantum chemistry simulations~\cite{Guseynov2022}, highlighting the value of gate diversity in optimizing quantum circuits.

However, direct implementation of the iSWAP gate remains difficult in quantum dot systems. Despite the fact that large qubit frequency differences are key to high-quality single-qubit operations, achieving strong exchange interactions requires operation far from the symmetric operating point (SOP), where the interaction is first-order immune to electric field fluctuations~\cite{martins2016noise,reed2016reduced}. In resonant iSWAP gate schemes~\cite{sigillito2019coherent, nguyen2023quantum}, the exchange interaction is ac modulated, with the frequency to be resonant with the difference in qubit frequencies. However, the non-linear relationship~\cite{van2018automated,cerfontaine2020high,pan2020resonant} between the barrier voltage and the exchange interaction makes it difficult to accurately shape the slope of the exchange interaction. Furthermore, the positivity of the exchange interaction necessitates an additional dc bias. This bias creates a persistent ZZ coupling that generates the control phase evolution concurrently with the ac-driven iSWAP operation, which requires precise calibration of the amplitude ratio for the ac and dc drives
~\cite{nguyen2023quantum}.

In addition, qubit coherence is a precious resource for quantum information processings, as it limits the performance of quantum operations. Although spin relaxation times $T_1$ can reach the millisecond level~\cite{RevModPhys.95.025003,ciriano2021spin,mills2022high} for semiconductor spin qubits, their pure dephasing times $T_2^*$ are limited to the microsecond level. Even with the help of the Hahn echo \cite{noiri2022fast,xue2022quantum,tanttu2024assessment}, the dephasing time $T_2^{\text{echo}}$ can only be improved to tens of microseconds, as it is limited by low-frequency noise~\cite{nakajima2020coherence,elman2017long}. Potential sources of such noise include hyperfine interactions from surrounding nuclear spins and charge noise~\cite{chan2018assessment,bermeister2014charge}. Although the isotopic purification technique can mitigate nuclear spin fluctuations~\cite{itoh2014isotope,eng2015isotopically}, low-frequency charge noise remains problematic, as it couples to external magnetic field gradients~\cite{mills2022two}, which is necessary for qubit operation. Therefore, suppressing the interference effect from low-frequency noise during gate operations remains imperative.

Here, we introduce a robust scheme for microwave-driven iSWAP gates on exchange-coupled spin qubits that uses single-qubit drives to continuously decouple the system from low-frequency noise during gate operation. To clearly present our scheme, we divide it into three progressive levels: a) a single-step scheme that, while effective, requires specific ratios of drive-exchange strength; b) a composite sequence that can eliminate the above constraint; and c) a dynamically corrected sequence~\cite{zeng2018fastest} providing first-order immunity to microwave amplitude errors. Numerical simulations verify that our scheme can achieve high fidelity with strong robustness under current experimental conditions, and thus provides a promising building block for scalable quantum computation.

\begin{figure}
    \centering
\includegraphics[width=0.9\columnwidth]{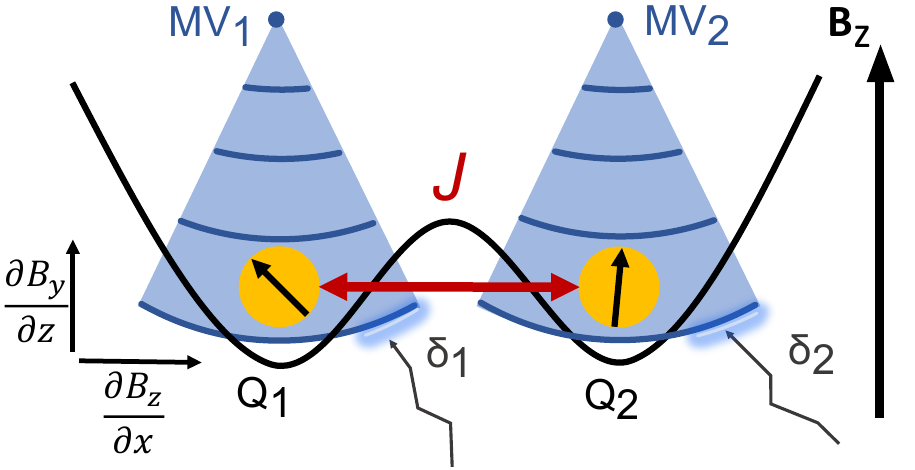}
\caption{Illustration of the microwave-driven iSWAP gate scheme. The two qubits $Q_1$ and $Q_2$ are coupled via an exchange interaction $J$. A uniform external magnetic field ($B_z$) provides the Zeeman splitting, while a local micromagnet creates an inhomogeneous field.
The resulting longitudinal field gradient ($\partial B_z/\partial x$)  ensures qubit addressability, while the transverse gradient ($\partial B_y/\partial z$)  facilitates EDSR.
Local EDSR microwave-drives $MV_1$ and $MV_2$ (shown as smooth blue lines) are used to drive the qubits. These drive fields simultaneously provide continuous dynamical decoupling from low-frequency noise $\delta_1$ and $\delta_2$ (shown as jagged grey lines).}
\label{fig1}
\end{figure}

\section{The Two-Qubit System}
\label{sec:2}
The two-qubit system under consideration comprises two gate-defined quantum dots, each confining a single electron. As shown in Fig.~\ref{fig1}, a uniform external magnetic field induces Zeeman splittings, while a micromagnet provides an inhomogeneous magnetic field for individual qubit addressing and electric dipole spin resonance (EDSR)~\cite{nowack2007coherent,pioro2008electrically,yoneda2018quantum,zajac2018resonantly}. The qubits are coupled via a tunable exchange interaction $J$, controlled by a barrier gate voltage while maintaining the double quantum dot (DQD) detuning in the charge-noise-immune SOP~\cite{martins2016noise,reed2016reduced}. The system is described by the Heisenberg exchange interacting Hamiltonian~\cite{loss1998quantum}:
\begin{equation}  H=J\mathbf{S}_1\cdot\mathbf{S}_2+\mathbf{S}_1\cdot\mathbf{B}_1+\mathbf{S}_2\cdot\mathbf{B}_2,
\end{equation}
where $\mathbf{S}_{k}=(S_x^k,S_y^k,S_z^k)^T$ with $k\in\{1,2\}$ are the spin operators of the two electrons, and we assume $\hbar = 1$ hereafter. The magnetic fields (in energy units) are $\mathbf{B}_k = (0, B_y^k(t), B_z^k)^T$, where $B_z^k$ represents the Zeeman energy and $B_y^k(t)$ describes the EDSR drives for single-qubit rotations. We focus on the weak-exchange regime ($J \ll \left|\Delta E_z\right|$), with $\Delta E_z=B_z^1-B_z^2$ being the Zeeman energy difference for the DQD, to maintain large Zeeman differences to address while staying near the SOP.
The primary source of decoherence for this quantum system is the low-frequency noise, which is modeled as $H_\delta=\sum_{k=1,2}\delta_{k}S_{z}^{k}$, with $\delta_k$ being the random frequency drift, and can be assumed to be quasi-static during gate operations~\cite{nakajima2020coherence}. Then, the complete system is described by $H_{\text{tot}}=H+H_\delta$.

We first briefly review the conventional implementation of the iSWAP gate with resonant ac modulation of $J$. That is, to implement an iSWAP gate on the DQD system with a large Zeeman difference for the two dots, the exchange interaction is modulated as $J(t) = J_{\mathrm{dc}} + J_{\mathrm{ac}}\cos(\omega t)$, with the modulation frequency chosen to be resonant with the Zeeman difference frequency~\cite{nguyen2023quantum}, i.e., $\omega=\left|\Delta E_z\right|$. The dc bias $J_{\mathrm{dc}}$ originates from the positivity of the exchange interaction, and single-qubit drives are inactive in this scheme. Under the rotating-wave approximation (RWA) and in the interaction picture, see Appendix~\ref{Appendix:A} for details, the exchange Hamiltonian reduces to
\begin{equation}
\label{eq:Hconv}
H_{\mathrm{conv}} = \sum_{k =1, 2} \delta_k S_z^k + J_{\mathrm{dc}} S_z^1 S_z^2 + \frac{J_{\mathrm{ac}}}{2} (S_x^1 S_x^2 + S_y^1 S_y^2),
\end{equation}
where the $J_{\mathrm{ac}}$ term mediates the iSWAP transition while the $J_{\mathrm{dc}}$ term induces the accumulation of the CPhase. An iSWAP gate requires complete population transfer in the antiparallel spin subspace $J_{\mathrm{ac}}T/2 = \pi$, and synchronization of the CPhase and iSWAP operations $J_{\mathrm{dc}} / J_{\mathrm{ac}} \in \mathbb{Z}^+$.

\begin{figure}[tbp]
  \centering
\includegraphics[width=\columnwidth]{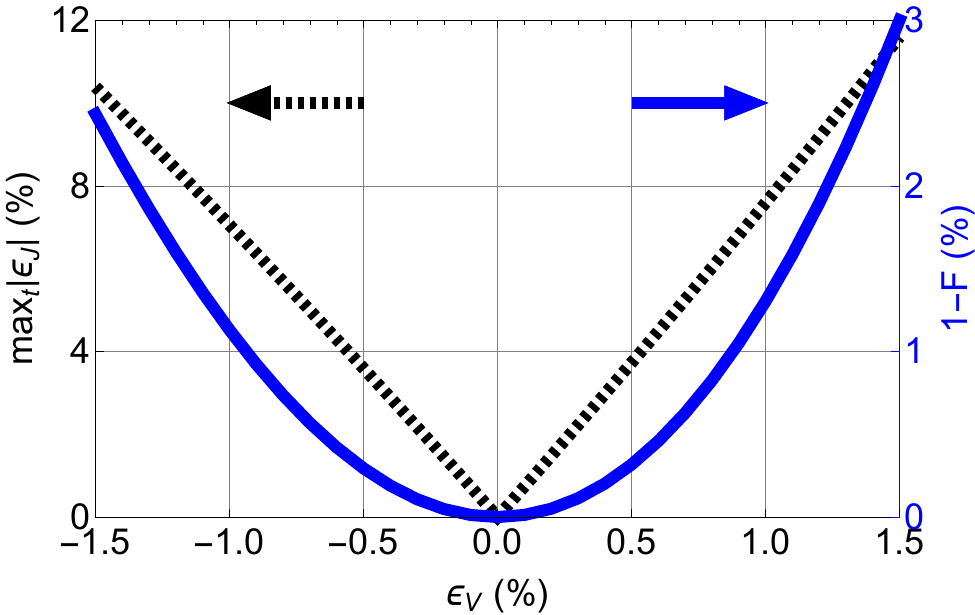}
\caption{Performance for the conventional resonant iSWAP gate scheme under control errors and decoherence. Gate infidelity ($1-F$, right y-axis) for the conventional scheme as a function of the relative error in the control voltage ($\epsilon_v$). The left y-axis shows the corresponding maximum relative error in the exchange interaction ($|\epsilon_J|_{\textrm{max}}$).
Numerical simulations include high-frequency noise ($T_2^{\textrm{echo}} = 20\,\mu\textrm{s}$) and low-frequency noise ($\sigma/2\pi = 0.1\,\textrm{MHz}$).}
  \label{fig2}
\end{figure}

However, low-frequency noise ($\delta_i$) directly limits the qubit coherence, and thus degrades the gate performance. Besides, the nonlinear dependence of the exchange interaction on the barrier voltage $v$ introduces critical control constraints. The typical relationship between $J$ and $v$ is $J(v) = J_0\exp(2\alpha v)$~\cite{van2018automated,pan2020resonant,cerfontaine2020high}, where $J_0$ is the residual exchange interaction and $\alpha$ is the lever arm. This implies that an experimental modulation $v(t) = v_0 + v_1 \cos(\omega t)$ produces:
$
J(t) = \exp(2\alpha v_0) \sum_{k=-\infty}^{\infty} I_k(2\alpha v_1) \exp(\mathrm{i} k \omega t)
$, where $I_k$ are modified Bessel functions of the first kind. This introduces an inherent dilemma: small driving amplitudes yield an insufficient $J_{\mathrm{ac}}$, extending gate times and exposure to decoherence; conversely, strong drives activate higher-order harmonics ($k \ge 2$) that perturb gate performance~\cite{nguyen2023quantum}. In addition, the requirement for precise control over the $J(t)$ profiles to realize the integer $J_{\mathrm{dc}}/J_{\mathrm{ac}}$ ratios further exacerbates the experimental challenges. Consequently, the conventional scheme faces a trade-off between gate speed and control complexity, rendering it vulnerable to both intrinsic noise and control imperfections.

To quantify the vulnerability of the conventional scheme to control imperfection in gate voltage, we introduce a relative voltage error fraction $\epsilon_v$, so that the applied voltage $v(t)$ is scaled to $v_\epsilon(t) = (1+\epsilon_v)v(t)$. In this scheme, the goal is to produce a perfect sinusoidal exchange interaction of $J(t) = J_{\mathrm{dc}} + J_{\mathrm{ac}}\cos(\omega t)$. The ideal voltage waveform required to achieve this is found by inverting the relation $J(v) = J_0 \exp(2\alpha v)$, which yields the form $v(t) = \ln [J_{\text{ideal}}(t)/J_0]/2\alpha$. The maximum relative error in $J$ is then taken over time $t$, that is, $\max_t |\epsilon_J|=\max_t |\{J[(1+\epsilon_v)v(t) - J[v(t)]\}/J[v(t)]|$. For the simulation in Fig.~\ref{fig2}, the operating point is defined by the target parameters $J_{\mathrm{dc}}/2\pi = J_{\mathrm{ac}}/2\pi = 7.5$ MHz, with $J_0/2\pi = 10$ kHz~\cite{mills2022two}. Notably, this simulation is independent of the lever arm $\alpha$, as it analytically cancels out. The error $\max_t |\epsilon_J|$ depends almost linearly on $|\epsilon_v|$ for small $\epsilon_v$, as the relative error is approximately $\max_t |\epsilon_J| \approx  |\epsilon_v|\max_t |2\alpha v(t)|$, a constant determined by the voltage profile. As shown in Fig.~\ref{fig2}, even a 1\% voltage offset leads to an approximately 10\% relative error in $J$, causing the gate fidelity to drop below 99\%, see Section~\ref{sec:4} for details.
Thus, it requires an alternative that is resistant to such control errors.

\section{Microwave-Driven iSWAP Gates}
\label{sec:3}
We present iSWAP gate implementations that utilize single-qubit microwave drives instead of ac modulation of the exchange interaction. This approach addresses both the susceptibility to low-frequency noise and the control challenges inherent in exchange interaction modulation. Our scheme achieves noise resilience by employing strong microwave drives to continuously decouple the system from low-frequency noise.
This simplifies the control of the exchange interaction from a complex and time-varying pulse shaping problem to a simple on/off switch. The dominant control error naturally shifts from imperfections in the gate voltage control to the microwave amplitude. Therefore, our scheme focuses on the robustness of the gate against low-frequency Zeeman noise and the control imperfection of the microwave amplitude.

\subsection{Scheme A: The direct implementation}
In our implementation, the driving microwave fields for qubits $k=1, 2$ are defined as $B_y^k(t) = 2\Omega_k\cos(\omega_k t + \phi_k)$, where $\Omega_k$, $\omega_k$, and $\phi_k$ denote the amplitudes, frequencies, and phases, respectively. For simplicity, we set the drive phases to $\phi_k = -\pi/2$. Under the RWA, which is valid when $B_z^k \gg \Omega_k$ and $\Delta E_z \gg J$, see Appendix~\ref{Appendix:A} for details, the total Hamiltonian in the interaction picture is
\begin{equation}
\label{eq:HtotRWA}
H_{\mathrm{RWA}}^{\mathrm{tot}} = \sum_{k=1,2} \delta_k S_z^k + \Omega_k S_x^k + J S_z^1 S_z^2.
\end{equation}
In the computational basis $\{\ket{\uparrow\uparrow}, \ket{\downarrow\uparrow}, \ket{\uparrow\downarrow}, \ket{\downarrow\downarrow}\}$, the Hamiltonian takes the matrix form of
\begin{equation}
    H_{\mathrm{RWA}}^{\mathrm{tot}} =
    \begin{pmatrix}
        \frac{\delta_1+\delta_2}{2} + \frac{J}{4} & \frac{\Omega_1}{2} & \frac{\Omega_2}{2} & 0 \\
        \frac{\Omega_1}{2} & \frac{\delta_2 - \delta_1}{2} - \frac{J}{4} & 0 & \frac{\Omega_2}{2} \\
        \frac{\Omega_2}{2} & 0 & \frac{\delta_1 - \delta_2}{2} - \frac{J}{4} & \frac{\Omega_1}{2} \\
        0 & \frac{\Omega_2}{2} & \frac{\Omega_1}{2} & -\frac{\delta_1+\delta_2}{2} + \frac{J}{4}
    \end{pmatrix}.
\end{equation}

Transforming the above Hamiltonian to the dressed-state basis $\ket{\pm}_k = (\ket{\uparrow}_k \pm \ket{\downarrow}_k)/\sqrt{2}$ via the unitary operator $S = \exp\left[ -\textrm{i} \pi (S_y^1 + S_y^2)/2 \right]$ yields
\begin{equation}
    H_{\mathrm{ds}}^{\mathrm{tot}} =
    \begin{pmatrix}
        \frac{\Omega_+}{2} & -\frac{\delta_1}{2} & -\frac{\delta_2}{2} & \frac{J}{4} \\
        -\frac{\delta_1}{2} & \frac{\Omega_-}{2} & \frac{J}{4} & -\frac{\delta_2}{2} \\
        -\frac{\delta_2}{2} & \frac{J}{4} & -\frac{\Omega_-}{2} & -\frac{\delta_1}{2} \\
        \frac{J}{4} & -\frac{\delta_2}{2} & -\frac{\delta_1}{2} & -\frac{\Omega_+}{2}
    \end{pmatrix},
\end{equation}
where $\Omega_\pm = \Omega_1 \pm \Omega_2$.
The noise-suppression mechanism here is analogous to continuous dynamical decoupling~\cite{Laucht2017,RevModPhys.88.041001,Stark2018}. The microwave drives $\Omega_k$ create an energy gap in the dressed-state basis $|\pm\rangle_k$, where the original noise term, $\delta_k S_z^k$, is transformed into a transverse perturbation, $-\delta_k S_x^k$. For  strong drivings satisfying $\Omega_k \gg |\delta_k|$, this perturbation is highly off-resonant with the dressed-state energy gap, suppressing its decoherence effects and thus shielding the gate dynamics. This noise suppression improves with increasing drive strength.

Under the strong driving condition of $\Omega_k \gg |\delta_k|$, the noise terms become negligible and the Hamiltonian simplifies to
\begin{equation}
H_{\mathrm{ds}} =
\begin{pmatrix}
\frac{\Omega_+}{2} & 0 & 0 & \frac{J}{4} \\
0 & \frac{\Omega_-}{2} & \frac{J}{4} & 0 \\
0 & \frac{J}{4} & -\frac{\Omega_-}{2} & 0 \\
\frac{J}{4} & 0 & 0 & -\frac{\Omega_+}{2}
\end{pmatrix}. \label{eq:dressed_hamiltonian}
\end{equation}
This partitions the Hilbert space into two decoupled subspaces: $S_1 = \operatorname{span}\{\ket{--}, \ket{++}\}$ and $S_2 = \operatorname{span}\{\ket{-+}, \ket{+-}\}$. With symmetric driving $\Omega_1 = \Omega_2 = \Omega$, the diagonal energy offset in the $S_2$ subspace, $\Omega_-$, vanishes. This enables resonant exchange oscillations in $S_2$ driven by the off-diagonal coupling $J/4$
\begin{equation}
\tilde{H}_{\mathrm{ds}} =
\underbrace{
\begin{pmatrix}
\Omega & \frac{J}{4} \\
\frac{J}{4} & -\Omega
\end{pmatrix}
}_{\text{$S_1$}}
\oplus
\underbrace{\begin{pmatrix}
0 & \frac{J}{4} \\
\frac{J}{4} & 0
\end{pmatrix}
}_{\text{$S_2$}}.
\end{equation}
The evolution in the $S_2$ subspace realizes an iSWAP gate at $T = 2\pi/J$. Meanwhile, the evolution in the $S_1$ subspace must be an identity up to a global phase, which requires phase matching satisfying $\sqrt{\Omega^2 + (J/4)^2} \cdot T = n \pi$ for $n\in \mathbb{N}_+$. This yields the drive-exchange relation of
\begin{equation} \label{eq:drive_condition}
\frac{\Omega}{J} = \frac{\sqrt{4n^2 - 1}}{4}.
\end{equation}
The resulting unitary operation is:
\begin{equation}
U_{\mathrm{direct}} = \begin{pmatrix}
(-1)^\textrm{n} & 0 & 0 & 0 \\
0 & 0 & -\textrm{i} & 0 \\
0 & -\textrm{i} & 0 & 0 \\
0 & 0 & 0 & (-1)^\textrm{n}
\end{pmatrix},
\end{equation}
which is equivalent to the iSWAP gate. However, this implementation imposes a strict requirement on the drive-exchange ratio, which demands precise experimental calibration.

\subsection{Scheme B: The composite scheme}
To remove the practical constraint on the drive-exchange ratio and enhance the gate's flexibility, we construct a composite gate sequence. The constraint originates from the need to synchronize the dynamics of the $S_1$ and $S_2$ subspace. In this scheme, we divide the total gate time $T = 2\pi/J$ into two equal segments and ensure that the $S_1$ subspace undergoes reciprocal dynamics, resulting in an identity operation. To achieve this cancelation, we introduce Z rotations defined on the dressed-state basis, $Z_{\phi}^k=\exp(-\textrm{i}\phi \tilde{\sigma}_z^k/2)$, where $\tilde{\sigma}_z^k$ is the Pauli-Z operator on the dressed-state basis. Experimentally, these rotations are realized by applying short microwave pulses, and contribute to the total gate time. The resulting transformation of the dressed-state Hamiltonian is
\begin{align}
\tilde{H}'_{\mathrm{ds}} &= Z_{-\pi/2}^1 Z_{-\pi/2}^2 \tilde{H}_{\mathrm{ds}} Z_{\pi/2}^1 Z_{\pi/2}^2 \nonumber \\
&= \begin{pmatrix}
\Omega & -J/4 \\ -J/4 & -\Omega
\end{pmatrix}
_{S_1} \oplus
\begin{pmatrix}
0 & J/4 \\ J/4 & 0
\end{pmatrix}
_{S_2}. \label{eq:transformed_hamiltonian}
\end{align}
The key insight here is that this transformation selectively flips the sign of the off-diagonal coupling in the $S_1$ subspace while leaving the $S_2$ subspace invariant.
The combining evolution of the composite sequence, with opposite driving amplitudes for the $S_1$ subspace, is
\begin{align}
U_{\mathrm{com}} & = Z_{-\pi/2}^1 Z_{-\pi/2}^2 U\left(-\Omega, \pi/J \right) Z_{\pi/2}^1 Z_{\pi/2}^2 U\left(\Omega, \pi/J\right)\notag\\
&=\begin{pmatrix}
1 & 0 & 0 & 0 \\
0 & 0 & -\textrm{i} & 0 \\
0 & -\textrm{i} & 0 & 0 \\
0 & 0 & 0 & 1
\end{pmatrix},
\end{align}
where $U(\Omega,t)=\exp\left[ -\textrm{i}\tilde{H}_{\mathrm{ds}}(\Omega) t \right]$. The evolution in the $S_2$ subspace yields a NOT operation, while the $S_1$ subspace undergoes an identity operation due to the sign reversal between segments. Thus, we implement a locally equivalent iSWAP gate for an arbitrary drive strength $\Omega$.

\subsection{Scheme C: The dynamically corrected scheme}
To address systematic amplitude errors $\Omega_k \rightarrow (1 + \epsilon_k) \Omega_k$, we further develop a dynamically corrected gate (DCG) sequence. We assume negligible errors in the single-qubit rotation gates due to their high fidelity and short duration. The evolution operator of the $S_1$ subspace in Scheme B is an identity composed of two mutually inverse dynamic processes, a feature that is not affected by amplitude errors. Therefore, based on our robust design of Scheme B, we focus on the $S_2$ subspace. Under symmetric driving ($\Omega_1 = \Omega_2 = \Omega$), the Hamiltonian with amplitude errors in the $S_2$ subspace is
\begin{equation}
H_{\epsilon}^{S_2} = \begin{pmatrix}
\epsilon \Omega & J/4 \\
J/4 & -\epsilon \Omega
\end{pmatrix},
\end{equation}
where $\epsilon=\epsilon_2-\epsilon_1$. We construct the following gate sequence
\begin{align}
U_{\mathrm{DCG}}^{S_2} = & \exp\left[-\textrm{i} H_{\epsilon}^{S_2}(-\Omega) \tau_1 \right] \exp\left[-\textrm{i} H_{\epsilon}^{S_2}(\Omega) \tau_2 \right] \nonumber \\
\times & \exp\left[-\textrm{i} H_{\epsilon}^{S_2}(-\Omega) \tau_2 \right] \exp\left[-\textrm{i} H_{\epsilon}^{S_2}(\Omega) \tau_1 \right],
\end{align}
where $\tau_{1,2}$ are operation times to be determined. Expanding $U_{\mathrm{DCG}}^{S_2}$ to the first order in $\epsilon$, we obtain
\begin{align}
U_{\mathrm{DCG}}^{S_2}  &=\begin{pmatrix}
 \cos{\left[\frac{J \tau}{2}\right]} & -\textrm{i}\sin{\left[\frac{J \tau}{2}\right]}  \\
 -\textrm{i}\sin{\left[\frac{J \tau}{2}\right]} & \cos{\left[\frac{J \tau}{2}\right]}
\end{pmatrix}\notag\\
& + \frac{4\Omega \epsilon}{J} \begin{pmatrix}
 0 &  EA   \\
 -EA  & 0
\end{pmatrix}
+O(\epsilon^2),
\end{align}
where $\tau=\tau_1+\tau_2$ and $EA= \cos\left(J\tau_2/2\right)-\cos\left(J\tau/4\right)^2 $ determines first-order errors, which cancel when $\tau_1 = \pi/(3J)$ and $\tau_2 = 2\pi/(3J)$. With these settings, the combined evolution is
\begin{align}
U_{\mathrm{DCG}}^{S_2}  &=\begin{pmatrix}
 0 & -\textrm{i} \\
 -\textrm{i} & 0
\end{pmatrix} +O(\epsilon^2).
\end{align}

\begin{figure}[tbp]
  \centering
\includegraphics[width=\columnwidth]{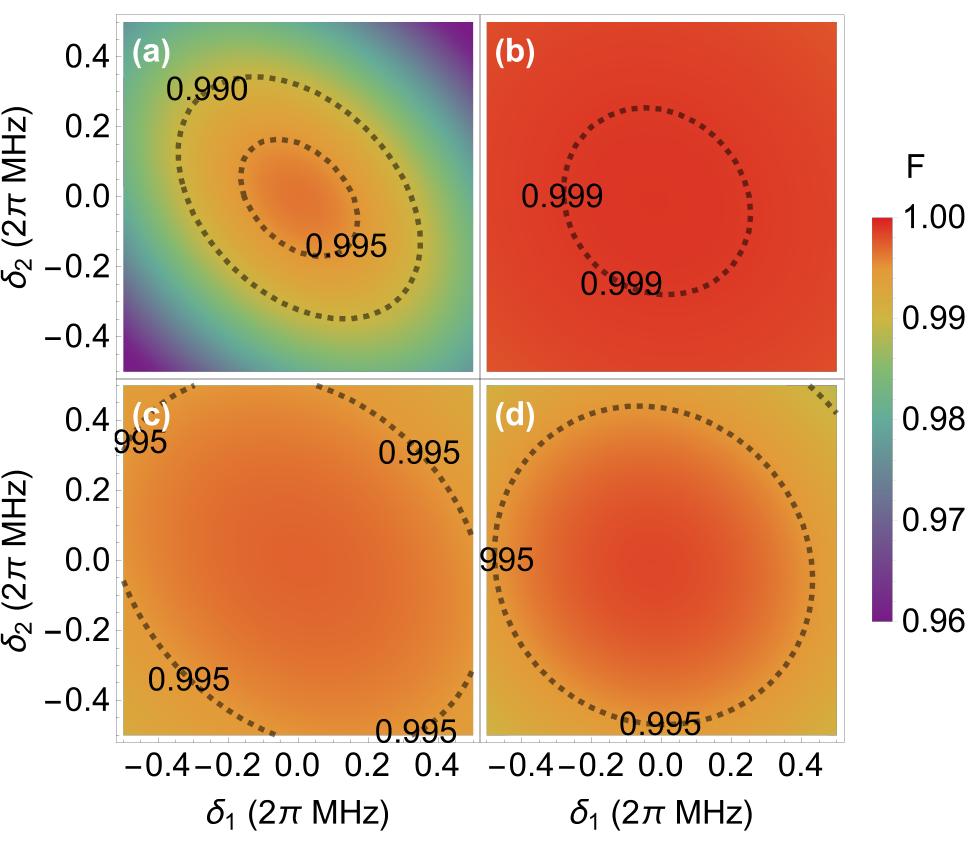}
\caption{Gate performance under decoherence.
The average gate fidelity $F$ as a function of low-frequency noise, modeled as quasi-static Zeeman shifts $\delta_{1,2}$. All simulations also account for the effects of high-frequency noise through a dephasing superoperator in the Lindblad master equation, with a rate corresponding to $T_2^{\text{echo}} = 20 \mu\text{s}$. The subplots compare the performance of (a) the conventional resonant iSWAP gate with the proposed (b) Scheme A, (c) Scheme B, and (d) Scheme C. The results demonstrate that our schemes are more robust, maintaining high fidelity across a wide parameter range,
coin contrast to the rapid fidelity degradation seen in the conventional scheme.}
  \label{fig4}
\end{figure}

\begin{figure*}[tbp]
  \centering
\includegraphics[width=2\columnwidth]{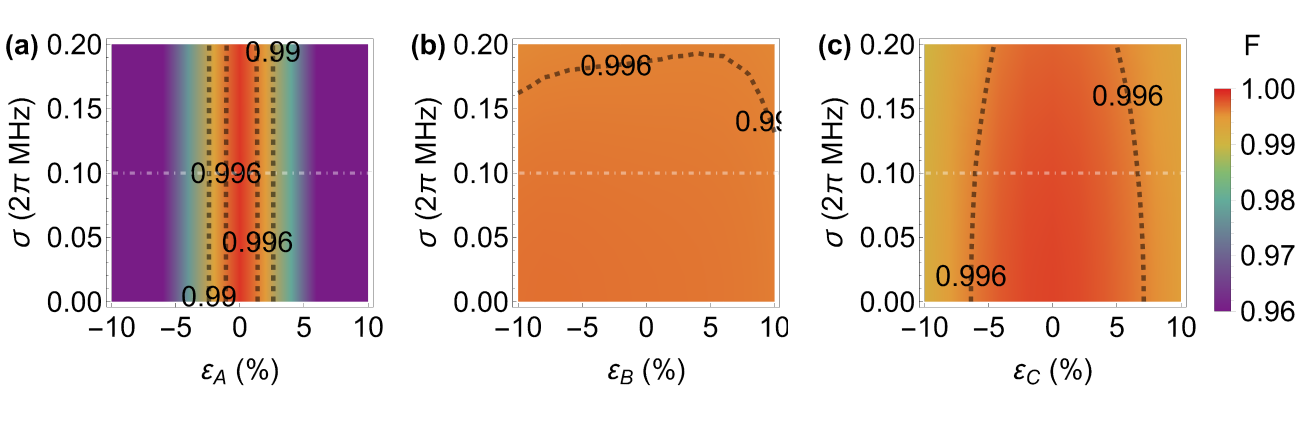}
 \caption{Gate fidelity F as a function of the low-frequency noise with standard deviation $\sigma$ and a symmetric amplitude error $\epsilon$, with $T_2^{\textrm{echo}} = 20\,\mu\textrm{s}$. The white dotted line at $\sigma/2\pi = 0.1\,$MHz indicates an experimentally relevant noise level. (a) Scheme A is sensitive to amplitude errors. Plots (b) and (c) correspond to Scheme B and Scheme C respectively, and shows their robustness against both noise sources simultaneously.}
 \label{fig:combined_noise}
\end{figure*}

The fully pulse sequence of the DCG scheme is constructed by replacing the two-qubit operation segments of Scheme B as $U(-\Omega, \pi/J) \rightarrow U\left[-\Omega,\pi/(3J)\right] U\left[\Omega,2\pi/(3J)\right]$ and $U(\Omega, \pi/J) \rightarrow U\left[-\Omega,2\pi/(3J)\right] U\left[\Omega,\pi/(3J)\right]$. This yields the complete sequence of
\begin{align}
U_{\mathrm{DCG}} = & Z_{-\pi/2}^1 Z_{-\pi/2}^2 U\left[-\Omega,\pi/(3J)\right] U\left[\Omega,2\pi/(3J)\right] \notag \\ \times & Z_{\pi/2}^1 Z_{\pi/2}^2
 U\left[-\Omega,2\pi/(3J)\right] U\left[\Omega,\pi/(3J)\right]\notag\\
 = & U_\text{com}+O(\epsilon^2),
\end{align}
which maintains noise suppression while providing first-order immunity to amplitude errors. Crucially, neglecting single-qubit gate times, this is achieved without extending the total gate-time, thereby offering enhanced robustness.

\section{Performance}
\label{sec:4}
To show the robustness of our scheme, we perform numerical simulation to evaluate the performance of our schemes under realistic conditions, including decoherence and control errors. The average gate fidelity $F$ was calculated by averaging the state fidelity $  F_{\rho_0} = \mathrm{tr}\left( \sqrt{ \sqrt{\rho_{T}} \rho_{\text{ideal}} \sqrt{\rho_{T}} } \right)$ over various input density matrices $\{\rho_0\}$,
where $\rho_{T}$ is the actual output density matrix and $\rho_{\text{ideal}}$ is the ideal output density matrix. The evolution of the system is modeled using the Lindblad master equation~\cite{lindblad1976generators,elman2017long} of
\begin{equation}
  \dot{\rho} = -\textrm{i} [H(t), \rho] + \sum_{k=1,2} \frac{\gamma_{k}}{2} \mathcal{L}(\sigma_z^k) \rho,
\end{equation}
where $H(t)$ is the Hamiltonian includes all counter-rotating terms. The superoperator is defined as $\mathcal{L}[c] \rho = c \rho c^{\dagger} -  c^{\dagger} c \rho/2 -  \rho c^{\dagger} c/2$ with $c=\sigma_z^k$ and $\gamma_{k}$ being the dephasing rate of qubit ${k}$. In our simulation, the spin relaxation effect is neglected, as $T_1$ is on the order of milliseconds, several orders of magnitude longer than our gate times (around 100 ns).

We used realistic experimental parameters~\cite{mills2022two}: the average Zeeman energy $E_{\text{avg}} = \left(B_z^1+B_z^2\right)/2 =  2\pi \times 17$ GHz, the Zeeman energy difference $\Delta E_z/2\pi = 0.3$ GHz, the maximum exchange interaction $J_{\text{max}}/2\pi = 15$ MHz, and the echo dephasing time $T_2^{\text{echo}} = 20 \mathrm{\mu s}$. For the conventional scheme, we set an ideal exchange pulse $J(t) = J_{\mathrm{dc}} + J_{\mathrm{ac}} \cos(\omega t)$, with $J_{\mathrm{dc}} = J_{\mathrm{ac}} = J_{\text{max}}/2$, to maximize the ac component $J_{\mathrm{ac}}$, implicitly assuming that this pulse profile could be generated from the nonlinear voltage response. In Scheme A, the microwave drive amplitude was set to $\Omega = (\sqrt{15}/4) J_{\text{max}}$ to satisfy the condition in Eq.~(\ref{eq:drive_condition}) for $n=2$.
Although schemes B and C do not require this specific ratio, we used the same $\Omega$ for a consistent comparison. The total gate time is about 66.7 ns for Scheme A and increases to 101.1 ns for Schemes B and C. The longer time is due to two additional Z-rotations. To ensure a realistic assessment, our simulations modeled these local rotations as physical microwave-driven gates, not as perfect operations, subject to the same decoherence and errors as the main entangling operation.

\begin{figure*}[tbp]
  \centering
\includegraphics[width=2\columnwidth]{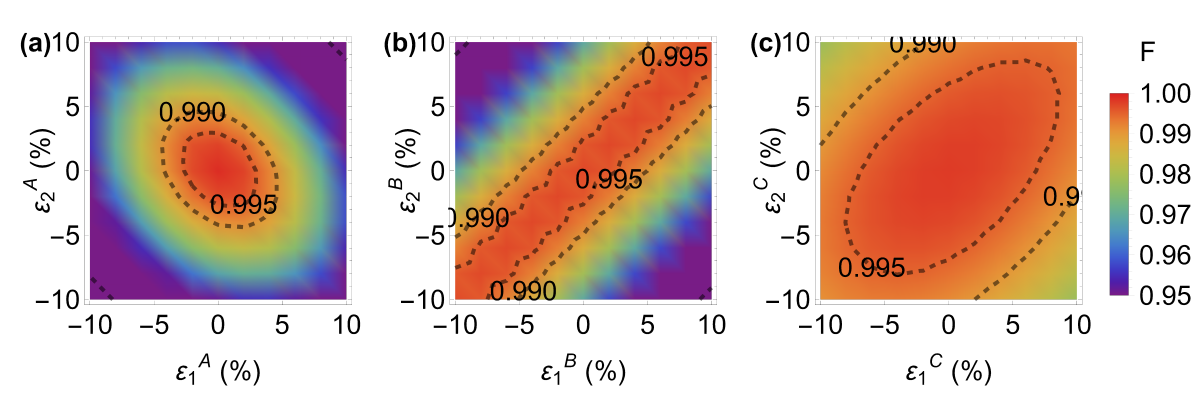}
\caption{Gate performance under control errors and decoherence. All simulations include high-frequency noise ($T_2^{\textrm{echo}} = 20\,\mu\textrm{s}$) and low-frequency noise ($\sigma/2\pi = 0.1\,\textrm{MHz}$). Plots (a), (b), and (c) show the gate fidelity ($F$) for the proposed Schemes A, B, and C, respectively, as a function of relative errors in the microwave drive amplitudes ($\epsilon_1, \epsilon_2$). The results clearly show the superior robustness of the dynamically corrected gate (Scheme C) against such errors, maintaining high fidelity across a wide parameter range.}
  \label{fig6}
\end{figure*}

Spin qubits are affected by both low-frequency and high-frequency noise~\cite{RevModPhys.95.025003,PhysRevX.10.011060,nakajima2020coherence,elman2017long}. In our simulations, low-frequency noise was modeled as quasi-static Zeeman shifts $\delta_i$, while high-frequency noise was incorporated through the dephasing superoperator in the Lindblad equation, corresponding to an echo dephasing time $T_2^{\text{echo}} = 20$ $\mu\text{s}$. Fig.~\ref{fig4} shows the gate fidelity under random Zeeman shifts $\delta_k / 2\pi \in [-0.5, 0.5]$ MHz.
The conventional scheme exhibits significant fidelity degradation as it is sensitive to low-frequency noise. In contrast, our schemes demonstrate substantially improved resilience, with the degree of noise suppression increasing with larger driving amplitudes.

In addition, our schemes achieve the iSWAP gate by simply switching the exchange interaction $J$ on or off, while leveraging microwave drives on individual qubits. This design eliminates the need for complex dynamic modulation of $J$, thereby avoiding the control challenges associated with its nonlinear behavior. Consequently, the critical control parameter shifts to the microwave drive amplitudes $\Omega_i$, which are generally easier to regulate. Therefore, we focused on testing the impact of microwave drive amplitude errors on the performance of our schemes. We investigated the effect of amplitude errors modeled as $\Omega_k \rightarrow (1 + \epsilon_k) \Omega_k$.

To analyze the performance under combined noise sources, we consider a scenario that includes both low-frequency noise and amplitude errors. We model the low-frequency noise on each qubit as an independent random variable drawn from the same Gaussian distribution, $\mathcal{N}(0, \sigma^2)$, and we assume a symmetric amplitude error ($\epsilon_1 = \epsilon_2 = \epsilon$). The results are shown in Fig.~\ref{fig:combined_noise}. The plot for Scheme A confirms it is resilient to the low-frequency noise but sensitive to amplitude errors. In contrast, Scheme B and C demonstrate strong robustness against both error sources simultaneously. The horizontal dotted line at $\sigma/2\pi = 0.1\,$MHz, which corresponds to an experimentally relevant pure dephasing time of $T_2^* \approx 2\,\mu\text{s}$~\cite{mills2022two}, based on the established relationship for Gaussian noise $T_2^* = \sqrt{2}/\sigma$~\cite{nakajima2020coherence}.

However, the assumption of perfectly symmetric errors ($\epsilon_1 = \epsilon_2$) may not hold in a real experimental setup. To assess our schemes in the most general case, we analyze their robustness to independent, asymmetric amplitude errors. Figs.~\ref{fig6} illustrate the gate fidelity under these errors.
These simulations include a realistic noise background, consisting of  high-frequency noise (characterized by $T_2^{\text{echo}} = 20\,\mu\text{s}$) and low-frequency noise drawn from a Gaussian distribution with a standard deviation of $\sigma/2\pi = 0.1\,\text{MHz}$.
As shown in Fig.~\ref{fig6}(a), scheme A shows some sensitivity to amplitude mismatches. As shown in Fig.~\ref{fig6}(b), while scheme B demonstrates enhanced robustness, its high performance is concentrated where errors are symmetric ($\epsilon_1 = \epsilon_2$). This vulnerability to error asymmetry, which cannot be guaranteed to be absent experimentally, is addressed by scheme C. As shown in Fig.~\ref{fig6}(c), scheme C maintains high fidelity over a broad range of amplitude errors, confirming its first-order robustness. Specifically, even with asymmetric amplitude errors of up to $\pm 5\%$, gate fidelity about 99.5\% can still be approximately obtained, which remains above the fault-tolerance threshold of $99\%$. Thus, scheme C shows a stark contrast to the rapid fidelity degradation in schemes A and B under similar asymmetric error conditions. While the composite schemes incur a longer gate time, this represents a favorable trade-off. The analysis demonstrates that this added complexity yields a substantial improvement in robustness against critical control errors.

\begin{figure}[tb]
 \centering \includegraphics[width=\columnwidth]{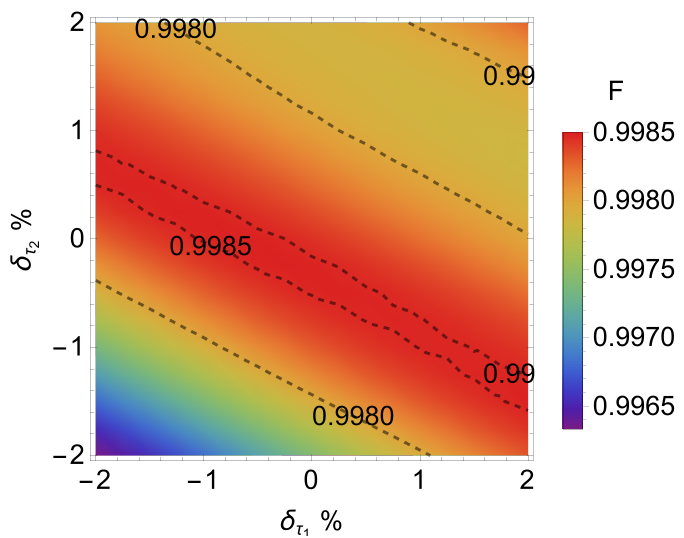}
 \caption{Gate fidelity of Scheme C as a function of independent relative timing errors $\delta_{\tau_1}$ and $\delta_{\tau_2}$ over a $\pm 2\%$ range, without decoherence. It shows strong robustness to timing errors over a wide parameter range.}
 \label{fig:timing_error}
\end{figure}

To specifically address the gate's sensitivity of scheme C to timing errors that would violate the ideal $\tau_2=2\tau_1$ ratio, we performed a decoherence-free numerical simulation. Given that modern arbitrary waveform generators offer sub-nanosecond timing precision, we analyzed the gate fidelity as a function of independent relative errors within a practical range of $\tau_1 \rightarrow \tau_1(1+\delta_{\tau_1})$ and $\tau_2 \rightarrow \tau_2(1+\delta_{\tau_2})$, where $|\delta_{\tau_1}|, |\delta_{\tau_2}| \le 2\%$. The results are shown in Fig.~\ref{fig:timing_error}. The plot demonstrates that our DCG scheme C is highly robust to this class of imperfections. Even for timing errors at the edges of this range, the fidelity remains consistently above 99.65\%, and exceeds 99.8\% over a wide parameter space, confirming the scheme's practical viability. And, the gate error here is mainly due to the RWA.

\section{Conclusions \label{sec:conclusion}}
In summary, we present a robust microwave-driven iSWAP gate for semiconductor spin qubits. Our scheme addresses two critical challenges, suppression of low-frequency noise and mitigation of control imperfections. Our scheme removes the need for complex ac modulation of the exchange interaction by simple, local microwave drives, which simultaneously provide continuous decoupling of qubits from their quasi-static noise. We have presented three progressively optimized protocols: a direct implementation (Scheme A) that demonstrates the core principle; a composite scheme (Scheme B) that removes operational constraints; and a DCG scheme (Scheme C) that provides first-order immunity to amplitude errors of the external microwave driving fields. Numerical simulations verify high fidelity and strong robustness of our scheme under realistic experimental parameters, diversifying the tool set available for quantum circuits.

Beyond semiconductor spin qubits, our dynamically corrected scheme could be adapted for other quantum systems governed by exchange-type interactions, such as certain superconducting qubit architectures. Future work could focus on the experimental implementation and characterization of this robust iSWAP gate, as well as exploring its integration into small-scale quantum algorithms to demonstrate its practical advantage in reducing circuit depth and mitigating errors.

\appendix

\section{Derivation of the Effective Hamiltonians}
\label{Appendix:A}

Here, we provide detail derivations for the effective Hamiltonians used in the main text. Both schemes can be understood as special cases of a general lab-frame Hamiltonian that includes the Zeeman interaction, quasi-static noise, and microwave drives. The exchange interaction is
\begin{equation}
  H_{\text{lab}} = H_0+ \sum_{k=1,2} \delta_k S_z^k +  B_y^k(t) S_y^k + J(t) \mathbf{S}_1 \cdot \mathbf{S}_2,
    \label{eq:H_lab_general}
\end{equation}
where $H_0 = \sum_k B_z^k S_z^k$ is the Zeeman terms. The analysis for both schemes proceeds by first transforming Eq. (\ref{eq:H_lab_general}) into an interaction picture, with respect to $H_0$, and  the Hamiltonian becomes
\begin{align}
H_{\text{int}} = & \sum_{k=1,2} \delta_k S_z^k + J(t) S_z^1 S_z^2\\ \nonumber
    & +\bigg[\frac{J(t)}{2}\textrm{e}^{-\textrm{i}\Delta E_z t}S_-^1 S_+^2+\sum_{k=1,2}\textrm{i}\frac{B_y^k(t)}{2} \textrm{e}^{-\frac{\textrm{i} B_z^k t}{2}}S_{-}^{k}+\text{H.c.} \bigg]
    \label{eq:H_lab}
\end{align}
where ladder operators $S_{\pm}^k = S_x^k \pm \textrm{i}S_y^k$ is the transverse part of the exchange interaction.

For the conventional scheme, the microwave drives are inactive ($B_y^k(t) = 0$), and the exchange interaction is modulated as $J(t) = J_{\mathrm{dc}} + J_{\mathrm{ac}}\cos(\omega t)$, with the modulation frequency $\omega = |\Delta E_z| = |B_z^1 - B_z^2|$. In the interaction picture,
\begin{align}
    H_{\textrm{int}}^{\textrm{conv}}(t) =
    & \sum_{k =1, 2} \delta_k S_z^k+
    [ J_{\mathrm{dc}} + J_{\mathrm{ac}}\cos(\Delta E_z t) ] S_z^1 S_z^2 + \\ \nonumber
    & \bigg[ \left(\frac{J_{\mathrm{ac}}}{4}+\frac{J_{\mathrm{dc}}}{2}\mathrm{e}^{-\mathrm{i}\Delta E_z t}+\frac{J_{\mathrm{ac}}}{4}\mathrm{e}^{-\mathrm{i}2\Delta E_z t}\right) S_-^1 S_+^2  + \text{H.c.}
    \bigg],
\end{align}
As $J_{\textrm{ac,dc}}\ll|\Delta E_z |$, we apply the RWA to neglect fast-oscillating terms, then we arrive the Hamiltonian in Eq.~(\ref{eq:Hconv}) of the maintext.

\begin{figure}[tb]

 \centering
 \includegraphics[width=\columnwidth]{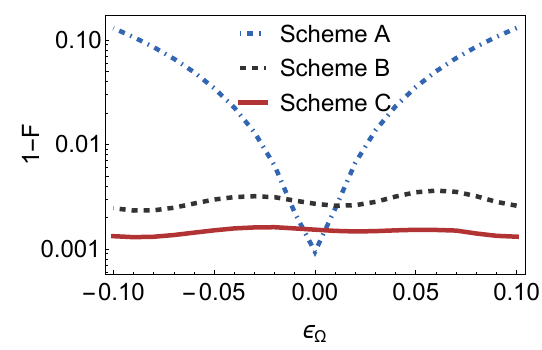}
 \caption{Gate infidelity ($1-F$) as a function of the relative drive strength deviation, $\epsilon_{\Omega} = 1 - \Omega/\Omega_{\text{ideal}}$, without decoherence. The fidelity of the ratio-dependent Scheme A is sensitive to the drive strength, while Schemes B and C are robust to such variations. The small, nearly constant infidelity in Schemes B and C is due to the RWA. The exchange strength is $J/2\pi=15$ MHz and the ideal drive strength for Scheme A is $\Omega_{\text{ideal}} = (\sqrt{15}/4)J$, and other parameters are the same as in Section \ref{sec:4}.}
 \label{fig:ratio}
\end{figure}

For our proposed microwave-driven scheme, the exchange interaction is a constant, $J(t) = J$, while the qubits are driven by resonant microwave fields, $B_y^k(t) = 2\Omega_k \cos(\omega_k t - \pi/2)$, with drive frequencies $\omega_k = B_z^k$. In the interaction picture with respect to $H_0$, we obtain
\begin{align}
    H_{\textrm{int}}^{\textrm{tot}}(t) =
    & \sum_{k =1, 2} \delta_k S_z^k+\bigg\{ \bigg[\frac{\Omega_k(t)}{2}+\frac{\Omega_k(t)}{2}\textrm{e}^{-2\textrm{i}B_z^k t}\bigg]S_{-}^{k}+\text{H.c.}
    \bigg\} \\ \nonumber
    &+ J S_z^1 S_z^2,
\end{align}
under the condition $\Omega_k\ll B_z^k$ and $J \ll |\Delta E_z|$, after the RWA, we obtian the effective Hamiltonian in Eq.~(\ref{eq:HtotRWA}) of the maintext.

\section{Robustness to Drive-Exchange Ratio Mismatch}
In Sec. III A, Scheme A requires a strict drive-exchange ratio as prescribed by Eq. ~(\ref{eq:drive_condition}). To explicitly quantify this challenge and to demonstrate the practical advantage of composite schemes B and C, we performed a decoherence-free numerical simulation comparing their performance. As shown in Fig.~\ref{fig:ratio}, the fidelity of Scheme A is critically dependent on this parameter, dropping sharply when the drive strength $\Omega$ deviates from the ideal condition. In contrast,  the composite pulse schemes B and C are, by design, insensitive to the drive strength. This simulation confirms that the composite pulse sequences effectively remove the strict parameter ratio requirement, offering a significant practical advantage.

\begin{figure}[tb]
 \centering
 \includegraphics[width=\columnwidth]{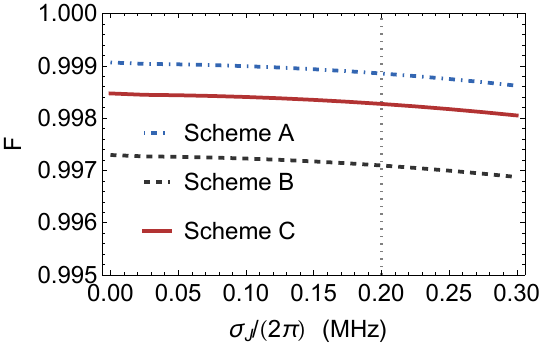}
 \caption{Robustness to exchange coupling noise. The simulation is decoherence-free. Average gate fidelity for Scheme A (blue, dot-dashed), Scheme B (black, dashed), and Scheme C (red, solid) as a function of the standard deviation, $\sigma_J$, of quasi-static J-noise. Each point is numerically averaged over the noise distribution. The vertical dotted line marks a realistic noise level of $\sigma_J/2\pi = 0.2$ MHz. The small, nearly constant infidelity in Schemes B and C is due to effects beyond the RWA. Other parameters are the same as in Section~\ref{sec:4}.}
 \label{fig:j_noise_appendix}
\end{figure}

\section{Fidelity under Quasi-static Exchange Noise}

In addition to the control errors discussed in the main text, we simulate the fidelity of the three proposed schemes under the exchange coupling noise of $J\rightarrow J+\delta J$. In our simulation, it is assumed that the fluctuation $\delta_J$ follows a Gaussian distribution with a mean of zero and a standard deviation of $\sigma_J$. The average gate fidelity was then calculated by numerically averaging the fidelity over this noise distribution.

The results are presented in Fig.~\ref{fig:j_noise_appendix}, which shows that all the three schemes exhibit high resilience to this noise. To establish a realistic noise level, we adopt singlet-triplet qubits, where dephasing is dominated by $J$-noise~\cite{cerfontaine2020high,PhysRevLett.124.117701}, with dephasing time of $T_2^* \approx 1 \mu\text{s}$ \cite{PhysRevLett.124.117701} corresponds to a standard deviation of $\sigma_J = \sqrt{2}/T_2^* \approx 2\pi \times 0.2$ MHz. In particular, at this level of noise, the fidelities for all schemes remain above 99.7\%. This shows that our proposed schemes are robust not only to control errors but also to intrinsic noise in the exchange coupling.

\acknowledgments
We thank Dr. Cheng-Xian Zhang for helpful discussions. This work was supported by the National Natural Science Foundation of China (Grant No. 12275090), the Guangdong Provincial Quantum Science Strategic Initiative (Grant No. GDZX2203001), and the Innovation Program for Quantum Science and Technology (Grant No. 2021ZD0302303).

%

\end{document}